\documentstyle[12pt, aaspp4]{article}

\slugcomment{25 July, 1998; To appear in ApJ (Main Journal)}

\begin{document}
\def\abs#1{\left| #1 \right|}

\def\EE#1{\times 10^{#1}}

\def\gcm{\rm ~g~cm^{-3}}
\def\cm3{\rm ~cm^{-3}}

\def\kms{\rm ~km~s^{-1}}
\def\cms{\rm ~cm~s^{-1}}

\def\ergs{\rm ~ergs~s^{-1}}

\def\isotope#1#2{\hbox{${}^{#1}\rm#2$}}
\def\wl{~\lambda}
\def\wll{~\lambda\lambda}
\def\HI{{\rm H\,I}}
\def\HII{{\rm H\,II}}
\def\HeI{{\rm He\,I}}
\def\HeII{{\rm He\,II}}
\def\HeIII{{\rm He\,III}}
\def\CI{{\rm C\,I}}
\def\CII{{\rm C\,II}}
\def\CIII{{\rm C\,III}}
\def\CIV{{\rm C\,IV}}
\def\NI{{\rm N\,I}}
\def\NII{{\rm N\,II}}
\def\NIII{{\rm N\,III}}
\def\NIV{{\rm N\,IV}}
\def\NV{{\rm N\,V}}
\def\NVI{{\rm N\,VI}}
\def\NVII{{\rm N\,VII}}
\def\OI{{\rm O\,I}}
\def\OII{{\rm O\,II}}
\def\OIII{{\rm O\,III}}
\def\OIV{{\rm O\,IV}}
\def\OV{{\rm O\,V}}
\def\OVI{{\rm O\,VI}}
\def\OVII{{\rm O\,VII}}
\def\OVIII{{\rm O\,VIII}}
\def\CaI{{\rm Ca\,I}}
\def\CaII{{\rm Ca\,II}}
\def\NeI{{\rm Ne\,I}}
\def\NeII{{\rm Ne\,II}}
\def\NeIII{{\rm Ne\,III}}
\def\NeIV{{\rm Ne\,IV}}
\def\NeV{{\rm Ne\,V}}
\def\NaI{{\rm Na\,I}}
\def\NaII{{\rm Na\,II}}
\def\NiI{{\rm Ni\,I}}
\def\NiII{{\rm Ni\,II}}
\def\FeI{{\rm Fe\,I}}
\def\FeII{{\rm Fe\,II}}
\def\FeIII{{\rm Fe\,III}}
\def\FeV{{\rm Fe\,V}}
\def\FeVII{{\rm Fe\,VII}}
\def\CoII{{\rm Co\,II}}
\def\CoIII{{\rm Co\,III}}
\def\ArI{{\rm Ar\,I}}
\def\MgI{{\rm Mg\,I}}
\def\MgII{{\rm Mg\,II}}
\def\SiI{{\rm Si\,I}}
\def\SiII{{\rm Si\,II}}
\def\SiIII{{\rm Si\,III}}
\def\SiIV{{\rm Si\,IV}}
\def\SiVI{{\rm Si\,VI}}
\def\SI{{\rm S\,I}}
\def\SII{{\rm S\,II}}
\def\SIII{{\rm S\,III}}
\def\SIV{{\rm S\,IV}}
\def\SVI{{\rm S\,VI}}
\def\FeI{{\rm Fe\,I}}
\def\FeII{{\rm Fe\,II}}
\def\FeIII{{\rm Fe\,III}}
\def\FeIV{{\rm Fe\,IV}}
\def\FeVII{{\rm Fe\,VII}}
\def\kI{{\rm k\,I}}
\def\kII{{\rm k\,II}}

\def\La{{\rm Ly}\alpha}
\def\Ha{{\rm H}\alpha}
\def\Hb{{\rm H}\beta}
\def\Lya{{\rm Ly}\alpha}

\def\etscale#1{e^{-t/#1^{\rm d}}}
\def\etscaleyr#1{e^{-t/#1\,{\rm yr}}}

\def\sigmaKN{\sigma_{\rm KN}}

\def\ncrit{n_{\rm crit}}

\def\Emax{E_{\rm max}}

\def\chieff{\chi_{\rm eff}^{\phantom{0}}}
\def\chieffi{\chi_{{\rm eff},i}^{\phantom{0}}}
\def\chiion{\chi_{\rm ion}^{\phantom{0}}}
\def\chiioni{\chi_{{\rm ion},i}^{\phantom{0}}}
\def\Gammaion{\Gamma_{\!\rm ion}}

\def\Mcore{M_{\rm core}}
\def\Rcore{R_{\rm core}}
\def\Vcore{V_{\rm core}}

\def\Menv{M_{\rm env}}
\def\Venv{V_{\rm env}}
\def\Vej{V_{\rm ej}}

\def\Vcthou{\left( {\Vcore \over 2000 \rm \,km\,s^{-1}} \right)}
\def\KK{\rm ~K}

\def\Msun{~M_\odot}
\def\Msunyr{~M_\odot~{\rm yr}^{-1}}
\def\Mdot{\dot M}
\def\Sdot{\dot S}

\def\tyr{t_{\rm yr}}
\def\gff{g_{\rm ff}}

\def\Tex{T_{\rm ex}}

\def\lsim{\!\!\!\phantom{\le}\smash{\buildrel{}\over
  {\lower2.5dd\hbox{$\buildrel{\lower2dd\hbox{$\displaystyle<$}}\over
                               \sim$}}}\,\,}

\def\gsim{\!\!\!\phantom{\ge}\smash{\buildrel{}\over
  {\lower2.5dd\hbox{$\buildrel{\lower2dd\hbox{$\displaystyle>$}}\over
                               \sim$}}}\,\,}

\title{Flash ionization of the partially ionized wind of the progenitor of SN 1987A.}

\author{Peter Lundqvist\altaffilmark{1}}

\altaffiltext{1}{Stockholm Observatory, SE-133 36 Saltsj{\"o}baden, Sweden; E-mail: peter@astro.su.se.}

\begin{abstract}
The H~II region created by the progenitor of SN 1987A was further heated and 
ionized by the supernova flash. Prior to the flash, the temperature of the gas 
was $\sim 4000 - 5000$~K, and helium was neutral, while the post-flash 
temperature was only slightly less than $\sim 10^5$~K, with the gas being 
ionized to helium-like ionization stages of C, N and O. We have followed the 
slow post-flash cooling and recombination of the gas, as well as its line 
emission, and find that the strongest lines should be N~V~$\lambda1240$ and
O~VI~$\lambda1034$. Both these lines are good probes for the density of the
gas, and suitable instruments to detect the lines are STIS on {\it HST} 
and {\it FUSE}, respectively. Other lines which may be detectable are 
N~IV]~$\lambda$1486 and [O~III]~$\lambda$5007, though they are expected to be
substantially weaker. The relative strength of the oxygen lines is found to be
a good tracer of the color temperature of the supernova flash. From previous 
observations, we put limits on the hydrogen density, $n_{\rm H}$, of the
H~II region. The early N~V~$\lambda1240$ flux measured by {\it IUE} gives
an upper limit which is $n_{\rm H} \sim 180~\eta^{-0.40} \cm3$, where $\eta$ 
is the filling factor of the gas. The recently reported emission in 
[O~III]~$\lambda$5007 at 2500 days 
requires $n_{\rm H} = (160\pm12)~\eta^{-0.19} \cm3$, for a supernova burst 
similar to that in the 500full1 model of Ensman \& Burrows (1992). For the more 
energetic 500full2 burst the density 
is $n_{\rm H} = (215\pm15)~\eta^{-0.19} \cm3$. These values are much higher
than in models of the X-ray emission from the supernova 
($n_{\rm H} \sim 75 \cm3$), and it seems plausible that the observed [O~III] 
emission is produced primarily elsewhere than in the H~II region. We also
discuss the type of progenitor consistent with the H~II region. In particular,
it seems unlikely that its spectral type was much earlier than B2 Ia.
\end{abstract}


\section{Introduction}
There is mounting evidence that the ejecta of supernova (SN) 1987A are 
interacting with a relatively dense ($\sim 10^2 \cm3$) circumstellar medium
(CSM). This CSM was first identified by Chevalier \& Dwarkadas (1995,
henceforth CD95) as an H~II region created by the supernova progenitor. 
Chevalier \& Dwarkadas based their model on the appearance of X-rays 
(Beuermann, Brandt, \& Pietsch 1994; Gorenstein, Hughes, \& Tucker 1994),
and reappearance and subsequent evolution of radio emission (Staveley-Smith 
et al. 1992; Ball et al. 1995; Staveley-Smith et al. 1996). Recently, Sonneborn 
et al. (1998) and Garnavich, Kirshner, \& Challis (1997b) have detected broad
($\sim 2.0\EE4 \kms$) Ly$\alpha$ and H$\alpha$ emission, respectively, using 
the Space Telescope Imaging Spectrograph (STIS) on the {\it Hubble Space 
Telescope} ({\it HST}), clearly indicative of ejecta/CSM interaction 
(Michael et al. 1998).

The model of CD95 for the formation of the CSM is related to the
interacting-winds model of Blondin \& Lundqvist (1993) and Martin \& Arnett
(1995), though CD95 also include the ionizing radiation from the progenitor.
The main difference compared to the earlier models is the existence of
a rather dense ($\sim 10^2 \cm3$), photoionized red supergiant wind inside 
the denser swept up shell of red supergiant wind. As in the previous models,
a large asymmetry of the density in the red supergiant wind seems necessary
for the inner ring observed around the supernova (e.g., Jakobsen et al. 1991)
to form. In the following we will refer to the photoionized red supergiant wind 
as the ``H~II region''.

The density estimated by CD95 for the H~II region was confirmed by Borkowski, 
Blondin, \& McCray (1997) to give a satisfactory emission of X-rays, as well
as to give a natural explanation to the presently slow expansion of the
radio-emitting region ($2800 \pm 400 \kms$ on day 3200, Gaensler et al. 1997); 
while the supernova ejecta may expand almost freely in the blue supergiant 
wind, they slow down to the observed expansion rate in the dense H~II region 
(CD95).

Further possible support to the model of CD95 is the diffuse [O~III] emission 
inside the inner ring reported by Wang et al. (1998). The level of this
emission is close to what Wang et al. claim should come from the H~II region 
in the model of CD95. Here we expand on the modeling of the line emission from
this gas. In particular, we focus on the UV emission lines and how they 
constrain the properties of the H~II region. This knowledge has implications 
for estimates of the type of progenitor and the properties of the blue 
supergiant wind inside the H~II region. Knowing the density structure inside
the inner ring helps to predict the ejecta/ring collision, and to constrain the
initial velocity of the ejecta.

In \S 2 we discuss the geometry and density of the H~II region, and in \S 3
the flash ionization of this gas. The line emission from the ionized gas is 
discussed in \S 4, and its implications in \S 5. We summarize our conclusions 
in \S6.

\section{Pre-flash properties of the H~II region}
CD95 discussed a 1-D model for the formation of the presupernova nebula. We 
have updated this model with recent observations to estimate the density
and temperature of the H~II region. Our model is then used in \S 3 to calculate 
the post-flash evolution of the nebula. For simplicity, we have maintained 
the 1-D scenario of CD95, though we comment on asphericity effects.

The H~II region is bounded in the equatorial plane on its outside by the inner 
ring. Assuming that the [O~I] image in Sonneborn et al. (1998) best displays
this boundary (see discussions in Lundqvist \& Sonneborn 1998, and Sonneborn 
et al. 1998), and that the distance to the supernova is 50 kpc, we find that 
the radius of this boundary is $R = 5.9\EE{17}$~cm. 

The inner edge of the H~II region can be estimated from the extent of the 
observed radio emission when this is revived. On day 1550, Gaensler et al. 
(1997) measure a radius of $\sim 0\farcs63$. Following Chevalier (1998), we 
assume that the radio turn-on occurs around 1150 days and that the ejecta then
have reached $\sim 0\farcs61$. This corresponds 
to $r_{\rm II} \sim 4.5\EE{17}$~cm, if we again assume 50 kpc for the distance.

With a density of $\sim 10^2 \cm3$ (CD95), the ionization balance in the
H~II region is nearly in steady state before the supernova flash. If we assume 
that He/H~$= 0.2$, as derived for the inner ring (Lundqvist et al. 1998), 
Case B recombination of hydrogen, and use the extent of the H~II region found
above, we can relate the density of the H~II region, $\rho_{\rm II}$, 
to the number of ionizing photons put out by the progenitor per 
second, $\Sdot$. If helium is not ionized by the progenitor (see below), we 
find

\begin{equation}
\rho_{\rm II} \approx 2.0\EE{-22}~\left({\Sdot - \Sdot_{\rm ring} \over 10^{45}~{\rm s}^{-1}}\right)^{0.5}~\left({T \over 5000~{\rm K}}\right)^{0.4} ~\gcm,
\end{equation}


\noindent where $T$ is the gas temperature. We have in Eqn. (1) taken into
account that some of the ionizing photons are consumed behind the ionization
front in the ring, rather than in the low-density H~II region. We have called
the rate of consumption of ionizing photons in the ring $\Sdot_{\rm ring}$.
A rough limit on $\Sdot_{\rm ring}$ can be obtained if one assumes that the 
ionization front is at a distance from the inside of the ring corresponding to 
optical depth unity at 13.6~eV for neutral gas. For a temperature of 5000~K 
and a hydrogen density of $10^4 \cm3$ in the ring this means
that $\Sdot_{\rm ring} \sim 3.2\EE{45}$~s$^{-1}$. However, as the newly ionized 
gas flows away from the ionization front its density decreases, which also 
decreases the value of $\Sdot_{\rm ring}$ needed to keep the gas ionized. While
we postpone a more detailed discussion on this to Mellema \& Lundqvist 1998 
[henceforth ML98], we note that Eqn. (1) provides an upper limit 
on $\rho_{\rm II}$ if we put $\Sdot_{\rm ring} = 0$.

It is thought that the progenitor was a B3~Ia star (Rousseau et al. 1978), and
such stars have $\Sdot \sim 3.7\EE{45}$~s$^{-1}$ (Panagia 1973). 
With He/H~$= 0.2$ we arrive at a number density of hydrogen 

\begin{equation}
n_{\rm H} \lesssim 68~\left({\Sdot \over 10^{45}~{\rm s}^{-1}}\right)^{0.5}~\left({T \over 5000~{\rm K}}\right)^{0.4} ~\cm3. 
\end{equation}

To estimate $T$ we have run photoionization models using a spectrum of a
B3~Ia star, aproximated by a diluted blackbody with an effective temperature of 
16,100~K and $\Sdot = 3.7\EE{45}$~s$^{-1}$ (Panagia 1973). Abundances are the 
same as in the calculations for the inner ring by Lundqvist et al. (1998), 
i.e., (C+N+O)/(H+He+Z) is $\approx 0.29$ times the solar ratio of Anders \&
Grevesse (1989), and N/C = 4.5 and N/O = 1.0. The atomic data are also the 
same. We find that the temperature in the H~II region is in the 
range $(4.0 - 4.5)\EE3$~K, and that helium is indeed neutral. 
As there is some uncertainty in the iron abundance of the gas (Borkowski et al. 
1997), we have also run a model with a factor of 2 depletion of iron so that 
Fe/H~$= 1.4\EE{-5}$. Fe~II cooling then becomes less important, and the 
temperature is slightly higher, $(4.5 - 5.0)\EE3$~K. It seems clear, however,
that $T$ must be substantially lower than the 8000~K assumed by CD95. This has
implications for models of the dynamics of the system (see \S 5.5). As a
typical value for $T$ we have adopted $T = 4500$~K. Inserting this into 
Eqn. (2), along with $\Sdot = 3.7\EE{45}$~s$^{-1}$, 
gives $n_{\rm H} \lesssim 125 \cm3$.

The values of $\Sdot$ in Panagia (1973) were derived for $T = 7000$~K. 
A lower value of $T$ gives a slightly lower value of $\Sdot$, 
so a better limit on $n_{\rm H}$ in our case is 
therefore $n_{\rm H} \lesssim 118 \cm3$. This limit is somewhat higher than the 
density used by Borkowski et al. (1997) ($150$~amu$\cm3$, 
i.e., $n_{\rm H} \sim 75 \cm3$ for He/H=0.20) to obtain good agreement 
between their modeled X-ray spectra and the data of Hasinger, 
Aschenbach, \& Tr\"umper (1996). We note, however, that we only obtain an
upper limit on $n_{\rm H}$, and that the value of Borkowski et al. (1997)
should be larger if the X-ray emitting gas is more concentrated to the 
equatorial plane than in their model. Our limit on $n_{\rm H}$, and the X-ray 
density are therefore consistent.

Guided by this discussion, we have chosen $n_{\rm H} = 120 \cm3$ as a 
standard density for the H~II region in our models in \S 3, though we have also
tested other densities. A higher density could be possible (see Eqn. [2]), if 
the progenitor was of earlier spectral type than B3~Ia. We note from Panagia 
(1973) that $\Sdot$ increases rapidly with earlier spectral type. The spectrum
also gets harder, which results in higher $T$. Both these effects permit
higher limit on $n_{\rm H}$ using Eqn. (2). For example, a B2~Ia
produces $\sim 4.1$ times more ionizing photons than a B3~Ia star, 
corresponding to a factor of $\gsim 2$ higher limit on $n_{\rm H}$.

Low-density cases are also important to investigate, not just because the X-ray
models indicate a low density and the fact that $\Sdot_{\rm ring}$ may be
non-negligible, but also because the density in the H~II region is expected to
decrease with distance away from the equatorial plane. How big this density
decrease might be is uncertain, but it is unlikely to be as large as in the
external red supergiant wind because the sound speed of the gas in the H~II 
region, $C_{\rm II}\sim~7.1~(T/5000~{\rm K})^{0.5} \kms$, is only marginally 
lower than the velocity of the inner ring ($v_{\rm r}\sim 10.3 \kms$; 
Crotts \& Heathcote 1992, Cumming \& Lundqvist 1997); overpressure in the 
equatorial plane of the H~II gas will create a transverse flow (Chevalier 1998) 
of nearly isothermal gas to partially compensate for the lower density in the
red supergiant wind toward the poles. It seems plausible that one has to go to
rather small polar angles to encounter a density which is a factor of $\sim 2$ 
lower than in the equatorial plane (see ML98 for further details). 

In our models in \S 3, the density therefore ranges 
between $n_{\rm H} = 60-240 \cm3$. Note that when the gas is fully ionized, 
the electron density is a factor $\approx 1.4$ higher than $n_{\rm H}$. 
The temperature of the gas just prior to the supernova flash was set to 
4500~K throughout the H~II region. The exact value is not important for our 
results. While hydrogen is ionized, helium is supposed to be neutral.

\section{Ionization of the H~II region by the supernova}
To simulate the ionizing radiation from the supernova outburst, we have used 
the 500full1 and 500full2 models of Ensman \& Burrows (1992). Both these 
bursts were used by Lundqvist \& Fransson (1996; henceforth LF96) to give good
fits to the light curves of the narrow UV lines from the ring (Sonneborn et al. 
1997), and they appear to be limiting cases to the actual burst of SN 1987A
(Blinnikov et al. 1998). The degree of ionization in the H~II region
is therefore similar to that in the innermost part of the ring in the models 
of LF96; in the case of 500full1, the dominant ionization stages of C, N and O
are C~V--VI, N~VI and O~VI--VII, while 500full2 with its higher number of 
ionizing photons ($\sim 2.2\EE{57}$ versus $\sim 1.0\EE{57}$) and peak 
color temperature ($\sim 1.4\EE6$ versus $\sim 1.1\EE6$~K) results in 
C~V--VI, N~VI--VII and O~VII.

The main difference of the H~II region compared to the inner ring is the 
temperature. While the ring is initially neutral, the ionized hydrogen in the 
H~II region cannot contribute to the heating of the gas. This was also 
discovered by Lundqvist \& Fransson (1991) when they calculated the 
post-explosion temperature of the preionized freely expanding blue supergiant 
wind. Thus, instead of having maximum temperatures of $\sim 1.6\EE5$ 
and $\sim 2.0\EE5$~K, as are the temperatures of the inner ring for 500full1 
and 500full2, respectively, the H~II region is in the case 
of $n_{\rm H} = 120 \cm3$ only heated to $(5.9-7.9)\EE4$~K 
and $(6.5-9.2)\EE4$~K, respectively, with the temperature increasing from the 
inner edge of the H~II region to the outer. For $n_{\rm H} = 60~[240] \cm3$,
the maximum temperatures are $\approx 7.3\EE4~[8.1\EE4]$~K 
and $\approx 8.2\EE4~[1.0\EE5]$~K, for 500full1 and 500full2, respectively.

The low density of the H~II region ensures slow cooling and recombination.
As a matter of fact, even in models with $n_{\rm H} = 240 \cm3$, the 
temperature decrease is only a few per cent
until day 1150, which is when we have assumed that the ejecta reach the 
H~II region. From this epoch, we have included the observed X-ray emission of 
the ejecta/wind interaction (Hasinger et al. 1996) approximated by, 

\begin{equation}
L_{\rm sh} = 1.36\EE{32}~{(t - 1150) \over 1850}~{\rm exp}(-E/440)~ \ergs~{\rm eV}^{-1}.
\end{equation}
\noindent Here $E$ is the photon energy in eV and $t$ the time in days since 
the explosion. While the X-ray emission is unimportant for the ionization of 
the ring (Lundqvist et al. 1998), it affects marginally the gas in the H~II 
region. For example, the difference in abundances of N~III and O~III for 
models with and without X-rays are of the order a few per cent at the present 
epoch ($t \sim 4250$~days), 
and ions of higher ionization stages are affected even less. 

A more important effect of the shock than providing ionizing photons is that 
its outward propagation decreases the fraction of the H~II region that can 
emit narrow lines. We have assumed that the shock velocity remains constant 
at $2800 \kms$ after it has reached the  H~II region. In the equatorial plane 
this means that the shock has now ($t \sim 4250$~days) 
reached $\sim 5.25\EE{17}$~cm. 

If we use the 500full1 burst, the temperature in the unshocked H~II region 
at the present epoch ($t \sim 4250$~days) is $\sim (7.0-7.5)\EE4$~K, regardless
of the density, while for 500full2, it is $\sim (7.7-8.2)\EE4$~K 
for $n_{\rm H} = 60 \cm3$ and $\sim (8.5-9.1)\EE4$~K for the higher
densities. For all these temperatures, the recombination time for N~VI to N~V 
is $\sim 2\EE4~(n_{\rm H} / 120 \cm3)^{-1}$~days (see Figure~1 of Lundqvist \&
Fransson, 1991), which means that 
only $\sim (15-20) \times (n_{\rm H}/120 \cm3)\%$ of N~VI has recombined 
to ionization stage V, or lower. A somewhat larger 
fraction of oxygen has recombined to the same ionization stages. However, 
almost no recombination proceeds to ionization stage II of either element
because of fast collisional ionization from stage II to III. This immediately
tells us that [O~II] and [N~II] lines must be observationally unimportant (see
also \S 4). The way the ionization depends on density, and the rather marginal 
difference in ionization for the two markedly different bursts we have used, 
shows that the burst spectrum is in general less important for the evolution 
of the ionization structure than the gas density.

\section{Line emission}
\subsection{Broad emission lines}
Borkowski et al.~(1997) showed that the interaction of the H~II region with
the blast wave should give rise to broad emission lines, especially
N~V~$\lambda$1240 and O~VI~$\lambda$1034. Although the lines are weaker
than the broad emission lines from the reverse shock, they will constitute an
increasing fraction of the total broad emission with time. The flux from the 
blast wave scales linearly with the preshock fraction of N and O in 
ionization stages N~I--V, $f_{\rm N}$, and O~I--VI, $f_{\rm O}$, respectively. 
For N~V~$\lambda$1240, Borkowski et al. adopted $f_{\rm N} = 0.08$ and for
O~VI~$\lambda$1034 they adopted $f_{\rm O} = 0.10$, considered to be lower 
limits at $\sim 10$ years. 

The value of $f_{\rm N}$ used by Borkowski et al. is similar to what we 
estimate in \S 3, assuming the gas is initially only in N~VI. Looking more
closely at our simulations, we note that $f_{\rm N}$ scales nearly linearly 
with density, as expected.
In particular for $n_{\rm H} = 120 \cm3$, $f_{\rm N} \sim 0.16~(0.11)$ for
500full1 (500full2). The lower value for 500full2 is due to a slightly higher 
temperature, and because $\sim 22$\% of the nitrogen starts out at N~VII 
instead of N~VI. The value of $f_{\rm N}$ adopted by Borkowski et al. (1997) 
therefore seems adequate for the density they used. The value we find 
for $f_{\rm O}$ at $\sim 10$ years is $\sim 0.44~(n_{\rm H}/120 \cm3)^{0.44}$ 
for 500full1 and $\sim 0.19~(n_{\rm H}/120 \cm3)^{0.83}$ for 500full2,
for the range of densities we have tested. This could mean a slightly higher
O~VI~$\lambda$1034 flux from the blast wave than in the models by Borkowski et
al. (1997).

\subsection{Narrow emission lines}
The slow evolution of the temperature and ionization of the H~II region has the
interesting effect that the UV lines that once dominated the emission from the 
inner ring (He~II~$\lambda1640$, C III]~$\lambda1909$, N~III]~$\lambda1750$, 
N~IV]~$\lambda1486$, N~V~$\lambda1240$ and O~III]~$\lambda1663$; cf. Sonneborn
et al. 1997), could still be emitted by the H~II region. This is displayed in 
Figures~1 (for 500full1) and 2 (for 500full2), where we show calculated fluxes
of He~II~$\lambda1640$, N~III]~$\lambda1750$, N~IV]~$\lambda1486$, 
N~V~$\lambda1240$ and O~III]~$\lambda1663$ for various densities. In the panels
showing N~V~$\lambda1240$ we have included the observations of Sonneborn et al.
(1997) since these are used in \S 5.3 to put limits on $n_{\rm H}$. The 
{\it International Ultraviolet Explorer} ({\it IUE}) observations of the other
lines are too close to the noise level (which is of the order of, or slightly 
less than $10^{-13}$~ergs~s$^{-1}$~cm$^{-2}$ for the dereddened fluxes; 
cf. LF96) to be useful for our analysis here. In the figures, we have also 
included the important lines [O~III]~$\lambda5007$ and O~VI~$\lambda1034$. 

Figures~1 and 2 are for two cases: one where we have taken into account the 
decrease of the volume of the unshocked H~II region with time, and another 
where this decrease has not been included. The latter could be more realistic 
if the H~II region is not spherically symmetric, but instead bulges out to 
large radii for small polar angles. 

If the density of the H~II region varies strongly with polar angle, 
a combination of models with different densities should be more realistic. 
The models in Figures~1 and 2 could be used for such an analysis, 
if the fluxes are rescaled according to the expected volumes of the different 
density components. The fluxes of all lines, except O~VI~$\lambda$1034, 
are extremely density-sensitive, which implies that the emission in a
multi-density environment will be heavily weighted toward the component 
with the highest density. Presumably, such a component would reside close 
to the equatorial plane.

Figures~1 and 2 show the same general behavior of the light curves, except 
for the O~VI line, which decays much faster in the 500full1 case. This is a 
result of the slightly higher initial ionization of oxygen in the 500full2 
case. This, in combination with an also slightly higher temperature 
in the case of 500full2, makes recombination to O~III proceed somewhat
slower than for 500full1. To display this more clearly, we show in Figure~3 
relative line fluxes at 4250 days, as functions of $n_{\rm H}$.
The reference line is N~V~$\lambda$1240.

As described in Figure 3, the displayed models are for four cases.
From these models, it is obvious that while the absolute fluxes of the lines 
differ markedly depending on whether the full, or reduced size of the H~II 
region is included (see Figures~1 and 2), the {\it ratios} of the lines are 
insensitive to the size of the emitting region (see Figure~3). The ratios are,
however, sensitive to the burst spectrum. In particular, the strengths of the 
oxygen lines relative to N~V~$\lambda$1240 at 4250 days differ strongly between 
the 500full1 and 500full2 cases, and there is also a small difference for the
He~II~$\lambda1640$/N~V~$\lambda$1240 ratio. 

We caution that Figures~1--3 display fluxes for the multiplets of the lines. 
In particular, the N~V and O~VI lines have two components, 
N~V~$\lambda\lambda$1238.82, 1242.80 and O~VI~$\lambda\lambda$1031.93, 1037.62, 
with the shortward components being twice as bright as the longward. The 
separation of the components corresponds to $\approx 9.62\EE2 \kms$ 
and $\approx 1.65\EE3 \kms$, respectively. For N~V~$\lambda$1240, the 
separation of its components should be large enough not to blend in STIS 
spectra.

The lines from the H~II region are intrinsically narrow. In the ring plane, the
velocity of the gas is expected to be lower than for the ring itself (see ML98).
It is in fact likely that thermal broadening of the lines are comparable to
the velocity broadening, especially for the He~II line, which has a thermal
broadening of $\sim 18 \kms$. In any case, the lines are significantly 
narrower than the emission from the hot spot or the ejecta/H~II region 
interaction, and should in STIS spectra really trace the spatial extent of the 
emission.

\section{Discussion}
\subsection{Density from existing observations}
Figures~1 and 2 show that the line emission from the H~II region is sensitive
to gas density. In particular, this is true for lines from ions of
ionization stages III and IV. It is therefore interesting that Wang et al. 
(1998) have reported a detection of diffuse [O~III]~$\lambda$5007 emission 
from the projected region inside the inner ring. The flux at 2500 days 
was $f_{\rm [O~III]} = (0.9\pm0.3)\EE{-14}$~ergs~s$^{-1}$~cm$^{-2}$. We have 
included their observation in Figures~1, 2 and 4, where Figure~4 shows
calculated [O~III]~$\lambda$5007 fluxes at 2500 days as functions 
of $n_{\rm H}$ for the same models as in Figure~3. It is evident
that the [O~III] flux is extremely sensitive to density 
(roughly as $f_{\rm [O~III]} \propto n_{\rm H}^{5.4}$), but that it is also 
somewhat sensitive to the burst spectrum. If we include the decrease of the 
H~II region due to the shock propagation, the calculated [O~III] flux agrees 
with the observed when $n_{\rm H} = (160\pm12)~\eta^{-0.19} \cm3$ for 500full1,
and $n_{\rm H} = (215\pm15)~\eta^{-0.19} \cm3$ for 500full2, where $\eta$ is
the filling factor of the gas. Because our models are for spherical symmetry
(cf. above), a low value of $\eta$ would apply if the gas is concentrated to 
the equatorial plane. 

The values we find for $n_{\rm H}$ from [O~III]~$\lambda$5007 are slightly 
higher than that of Wang et al. (1998). This is not surprising since Wang et 
al. only studied the 500full1 case, and did not include the decrease of the 
H~II region due to the shock propagation. In both our study and that of Wang 
et al., the estimated densities are upper limits. This is not just 
because $\eta$ could be less than unity, but also because part of the 
diffuse [O~III] emission could be produced elsewhere than in the H~II region, 
and that it is only when seen in projection that the emission appears to come 
from inside the inner ring. (Note that Wang et al. only detect and measure 
emission from inside the ring.)

Our upper limits on $n_{\rm H}$ from the [O~III] line may seem compatible with
the estimated density from X-ray models of the supernova (Borkowski et al. 
1997), $n_{\rm H} \sim 75 \cm3$ (for He/H = 0.20). However, the 
factor of $\gsim 2$ difference between the X-ray and [O~III] densities is
not so easy to understand. A possible explanation could be that the [O~III]
emission probes gas further out in the H~II region than that at the shock front,
and that the density further out is higher. A problem with this interpretation
is, however, that the observed [O~III] emission is diffuse and not concentrated 
toward the ring (Wang et al. 1998).
Another possibility could be that the spectrum of the burst was softer than in
500full1. This would give a lower value of $n_{\rm H}$ in the [O~III] models.
Again, this is rather unlikely because this would contradict the finding of
LF96 that the burst spectrum should be at least as hard 
as 500full1 in order to model the emission from the inner ring (see 
also \S 5.3). The successful modeling of the UV emission lines from the ring 
also rules out that poor atomic data could cause the different densities in 
X-ray and [O~III] models. (The same data were used both here, and by 
LF96.)

We are therefore left with the possibilities that the observed [O~III] emission
at 2500 days is primarily formed elsewhere than in the H~II region, or that the 
X-ray models somehow underestimate the density. If the X-ray density 
($n_{\rm H} \sim 75 \cm3$) is correct, 
Figure 4 shows that the expected [O~III] flux at 2500 days is of 
order $\sim 10^{-16}$~ergs~s$^{-1}$~cm$^{-2}$. This is far too low to be 
detected, and the detected [O~III] emission can not come from the H~II region
if the density is this low. Possible sites of contaminating emission could
be the outer rings, the walls of the bipolar nebula, and/or the ejecta/H~II 
region interaction region. It should be noted that the observations of Wang 
et al. were made through [O~III] filters which can not discriminate between 
broad and narrow emission lines.

The X-ray density could be higher than $n_{\rm H} \sim 75 \cm3$ if the
volume of the X-ray emitting gas is smaller than in the model of Borkowski
et al. (1997). This could actually be the case since Borkowski et al. assume 
a rather small value of $r_{\rm II}$ compared to ours, $3.3\EE{17}$~cm 
and $4.5\EE{17}$~cm, respectively. If a smaller X-ray emitting region
could allow for a density as high as $n_{\rm H} \sim 120 \cm3$, this could be
consistent with a B3 Ia nature of the progenitor (see \S 2). However, even 
for such a high density, the [O~III] flux at 2500 days would only be of 
order $\sim 10^{-15}$~ergs~s$^{-1}$~cm$^{-2}$. So, for this density too, 
[O~III] emission from elsewhere than the H~II region is needed to obtain the 
flux measured by Wang et al. (1998).

Wang et al. (1998) also give an upper limit to the H$\alpha$ flux at 2500 days.
Assuming a similar spatial distribution of H$\alpha$ compared to that of 
[O~III]~$\lambda$5007, their H$\alpha$ flux 
is $f_{\rm H\alpha} \lesssim 0.6\EE{-14}$~ergs~s$^{-1}$~cm$^{-2}$. In our most
favorable model for H$\alpha$ emission, this only corresponds to a 
limit $n_{\rm H} \gtrsim 220 \cm3$. H$\alpha$ is thus not a good probe 
for $n_{\rm H}$.

Finally, we comment on the flux of N~V~$\lambda$1240 measured by the {\it IUE}
at 1600 -- 1800 days. Assuming $E(\bv) = 0.16$, with $E(\bv) = 0.10$ from the
LMC and $E(\bv) = 0.06$ from the Galaxy, and using the extinction curves of
Savage \& Mathis (1979) and Fitzpatrick (1985), the dereddened 
flux is $\sim 2.5\EE{-13}$~ergs~s$^{-1}$~cm$^{-2}$, though there is a large
scatter in the data (see Figs.~1 and 2). Adopting this flux, our simulations 
put an upper limit on $n_{\rm H}$ from N~V~$\lambda$1240, which 
is $n_{\rm H} \sim 180~\eta^{-0.40} \cm3$ for 500full1, and nearly
the same for 500full2. This limit is close to what we find for the [O~III]
flux at 2500 days, which again indicates that the detected [O~III] emission 
does not mainly come from the H~II region.

\subsection{Which lines to observe and what they probe}
Obviously, more observations are needed to explain the discrepancy 
described in \S 5.1. If we assume that $\sim 10^{-14}$~ergs~s$^{-1}$~cm$^{-2}$
is a reasonable limit to detect [O~III]~$\lambda$5007 at 4250 days, 
the density has to be $\gsim 130-140 \cm3$ for a burst like 500full1 to give
rise to detectable [O~III] emission. Although, such a density is on the high 
side compared to X-ray models, it can not be ruled out from our present 
knowledge. 

To probe lower values of $n_{\rm H}$ one should focus on the N~IV], N~V and 
O~VI lines. They have high excitation energies, and are therefore less
confused with emission from the low-temperature inner ring than, e.g., 
[O~III]~$\lambda$5007, which has had a rather constant temperature 
of $\sim 2.5\EE4$~K since $\sim 1200$~days (Lundqvist et al. 1998).

There is some contamination of N~V~$\lambda$1240 and O~VI~$\lambda$1034
from the shocked gas in the ejecta/wind interaction zone
(Borkowski et al. 1997). This diffuse emission is characterized by 
a range of velocities from a few hundred$\kms$ (in possible hot spots) up 
to $\sim 20,000 \kms$ (Sonneborn et al. 1998). To separate out those emissions 
from the narrow lines from the H~II region, it is essential that the spectral
resolution is $\lesssim 100 \kms$. This is achieved with both the
{\it Far-Ultraviolet Spectroscopic Explorer} ({\it FUSE)} 
for O~VI~$\lambda$1034, and {\it HST}/STIS for the rest of the lines.

At a density of $n_{\rm H} = 120 \cm3$ and for 500full1, the expected fluxes 
of the O~VI, N~V and N~IV] lines are $\sim 1\EE{-13}$, $\sim 1\EE{-13}$ 
and $\sim 4\EE{-14}$~ergs~s$^{-1}$~cm$^{-2}$, respectively. Including
reddening, which introduces the correction factors $\sim 0.11$, $\sim 0.17$ 
and $\sim 0.24$, respectively, for $E(\bv) = 0.16$ (cf.~above), and the 
possibility that $\eta < 1$, the lines may still be observable. 
For 500full2, the fluxes of the O~VI and N~V lines are
similar to those in models using 500full1, whereas N~IV] is down by a 
factor of $\sim 2$ (cf. Figs.~1 and 2). Lines from ionization stage III will
not be observable for any of the bursts, unless the density is higher. However, 
if it is, these lines will be extremely useful, since, e.g., the O~VI/[O~III] 
and O~VI/O~III] line ratios are sensitive to the burst spectrum. At a density 
of $n_{\rm H} = 75 \cm3$, the prospects of detecting any lines but the 
O~VI and N~V lines seem futile. 

Although the UV lines are the most useful to observe in order to narrow down
the value of $n_{\rm H}$, limits on [O~III]~$\lambda$5007 are certainly
useful too. Using STIS at the present epoch when the [O~III] 
emission from the H~II region should be stronger than at 2500 days, means
that [O~III] can probe lower densities than at 2500 days. As stated above, 
the [O~III] line may also be useful to place limits on the burst spectrum 
when combined with the O~VI line.

\subsection{Implications for the early UV line emission}
LF96 found that a successful modeling of the early UV lines from the inner
ring requires that the inner part of the ring be ionized to at least N~VI. 
Obviously, this means that also the H~II region should be ionized to N~VI, 
or higher. While this in accord with our results in \S 3, there are
possible indications from the early UV line emission studied by {\it IUE}
(Sonneborn et al. 1997) that the ionization of the H~II region could have
been lower. 

The {\it IUE} observations showed that the N~V~$\lambda$1240 
flux turned on between $70.2 - 80.6$ days, and that N~IV]~$\lambda$1486 turned
on a few days later, some time between $80.6 - 85.2$ days. That this is just
a recombination effect in the ring is not supported by the results of LF96. 
It is more likely that the early N~V~$\lambda$1240 emission comes from a 
separate region which has a shorter light travel time than the near side of 
the ring, and that this gas is devoid of N~IV. Possible such sites could be 
the H~II region, or a part of the bipolar nebula. 

If the early N~V~$\lambda$1240 emission comes from the H~II region, the
emitting gas can certainly not be spherically symmetric, since {\it IUE} would 
otherwise have detected N~V~$\lambda$1240 much earlier than at 80.6 days. 
To be consistent with the observations, the part of the H~II region possibly
emitting N~V~$\lambda$1240 should instead be concentrated toward the ring, 
with an inner edge of its emission at $\gtrsim 0.82 R$, assuming that 
N~IV]~$\lambda$1486 only comes from the inner ring. This radius is not very 
different from our assumed inner radius of the H~II 
region, $r_{\rm II}/R \approx 0.76$. The small separation in time between
the turn-on of N~V~$\lambda$1240 and N~IV]~$\lambda$1486 is therefore not 
enough to rule out that the $\sim 10^2 \cm3$ gas in H~II region could be 
responsible for the early N~V~$\lambda$1240 emission. 

It is, however, difficult to tie such a scenario into our model, especially
since we find that nitrogen is ionized to N~VI--VII throughout the H~II region
even for densities substantially higher than $240 \cm3$, and we have already
ruled out such high densities as the general density of the H~II region on 
several grounds (cf. above). A possibility is that the tentative high-density 
region emitting early N~V~$\lambda$1240 could be similar to the inward 
protrusion observed by Garnavich, Kirshner, \& Challis (1997a) and Sonneborn 
et al. (1998). If this is correct, the protrusion should be ionized beyond 
N~IV from the outset. Such a protrusion should reveal itself within the next
few years as a hot spot on the near side of the ring. The alternative scenario
for the early N~V~$\lambda$1240 emission is that the line forms in the 
presumably existing bipolar nebula. Again, this gas should not contain any 
N~III or N~IV zones giving rise to detectable emission from the outset.

Another possible indication of early N~V in the H~II region is that
resonance scattering in the H~II region by N~V~$\lambda\lambda$1238.82, 
1240.80 may be necessary to explain the somewhat delayed peak of the 
N~V~$\lambda$1240 light curve compared to the peak of the other UV lines
(Sonneborn et al.~1997). However, skipping the single high data point for 
N~V~$\lambda$1240 at $t = 485$ days, LF96 showed that the expected scattering
in the bipolar nebula is sufficient to delay the N~V~$\lambda$1240 emission. 
To check whether the H~II region could add additional scattering, we have 
calculated the column density of N~V through the H~II region in our models. 
We find that the column density is too low ($\lesssim 10^{11}$ cm$^{-2}$) in 
all models to cause an optical depth unity in N~V~$\lambda\lambda$1238.82, 
1240.80.

While the column density of the H~II region in our spherically symmetric
models is low, a larger column density may be obtained along the line
of sight to the rear side of the ring if the H~II region bulges out far from
the equatorial plane. The line of sight to the rear side of the ring could
then pass through geometrically thick parts of the H~II region on the near
side of the nebula which may cause resonance scattering. However, the velocity
shift between these two regions will be too large to result in any important
resonance scattering. We therefore conclude that neither the early emergence 
of N~V~$\lambda$1240, nor its subsequently delayed emission, contradict the 
idea that the H~II region was ionized to at least N~VI in the equatorial plane
(see \S 3.). This is also in accord with the idea that the inner side of the 
inner ring is ionized to N~VI (LF96).

\subsection{Implications for the progenitor}
The value of $n_{\rm H}$ for the H~II region has implications for the 
nature of the progenitor through Eqns.~(1 \& 2). Even with the high value 
of $n_{\rm H}$ we find from the [O~III] flux at 2500 days (see \S 5.1), the 
number of ionizing photons is

\begin{equation}
\Sdot \lesssim \Sdot_{\rm ring} + 5.5~ (10.0)\times 10^{45}~\left({T \over 5000~{\rm K}}\right)^{-0.8}~\eta^{-0.38}~{\rm s}^{-1} 
\end{equation}

\noindent for 500full1 (500full2). If $\eta$ is not $\ll 1$, 
and $\Sdot_{\rm ring}$ is not larger than estimated in \S 2, a progenitor more
powerful than B2 Ia seems to be ruled out. On the other hand, if we take
the density from the X-ray modeling to be a lower limit on $n_{\rm H}$, we 
find $\Sdot \gtrsim 1.2\EE{45}~(T/5000~{\rm K})^{-0.8}$~s$^{-1}$. This excludes
a progenitor much less powerful than a B3 Ia, which 
has $\Sdot \sim 3.7\EE{45}$~s$^{-1}$ (Panagia 1973).
The most likely range of spectral types of the progenitor therefore appears to
be B4 Ia -- B2 Ia, assuming a supergiant classification, which is fully 
consistent with the classification of Rousseau et al. (1978). A caveat is, 
however, that multidimensional effects could require a lower photoevaporation 
rate of the inner ring than in the 1--D case (see \S 5.6), which could push 
down the ``lower limit'' of the spectral type to less ionizing than B4 Ia.

\subsection{Implications for the blue supergiant wind}
From our model of the H~II region, we can use pressure balance arguments to
sketch the structure of the blue supergiant wind interior to the H~II region.
In a 1--D scenario, pressure balance between the shocked blue supergiant wind 
and the H~II region
gives $\rho_{\rm II}~C_{\rm II}^2 = 0.88~\rho_{\rm b}~u_{\rm b}^2$,
where $\rho_{\rm b}$ and $u_{\rm b}$ are the density and velocity of the
freely expanding blue supergiant wind at the termination shock, i.e., at a
radius $r_{\rm b}$. The factor 0.88 stems from Weaver et al. (1977; see also
CD95 and ML98). The density of the free wind is given by

\begin{equation}
\rho_{\rm b} \approx 1.0\EE{-25} \left({\Mdot_{\rm b} \over 
10^{-8} \Msunyr}\right) \left({u_{\rm b} \over 500 \kms}\right)^{-1}
\left({r_{\rm b} \over 10^{17}~{\rm cm}}\right)^{-2}~\gcm,
\end{equation}

\noindent where $\Mdot_{\rm b}$ is the mass loss rate of the blue supergiant
wind. From this we find that

\begin{equation}
\Mdot_{\rm b} \approx 4.7\EE{-9} \left({r_{\rm b} \over 10^{17}~{\rm cm}}\right)^{2} \left({T \over 5000~{\rm K}}\right)^{1.4}
\left({\Sdot - \Sdot_{\rm ring} \over 10^{45}~{\rm s}}\right)^{0.5} \left({u_{\rm b} \over 500 \kms}\right)^{-1} \Msunyr,
\end{equation}

\noindent which should be compared with the maximum value estimated by
Chevalier (1998) to be consistent with the fast expansion of nearly
undecelerated 
ejecta, $\Mdot_{\rm b} \lesssim 6.5\EE{-9}~(u_{\rm b}/500 \kms) \Msunyr$.
Eliminating $\Mdot_{\rm b}$ from these expressions, we find an upper limit
on $r_{\rm b}$, which is

\begin{equation}
r_{\rm b} \lesssim 1.2\EE{17}~\left({u_{\rm b} \over 500 \kms}\right)^{2}
\left({T \over 5000~{\rm K}}\right)^{-0.7}
\left({\Sdot - \Sdot_{\rm ring} \over 10^{45}~{\rm s}}\right)^{-0.25}~{\rm cm}.
\end{equation}

If we consider $u_{\rm b} = 750 \kms$ an upper limit on wind 
velocities of B3 Ia stars (e.g., Lamers, Snow, \& Lindholm 1995), and 
use $T = 4500$~K, we find that $r_{\rm b}$ cannot be larger 
than $\sim 1.4\EE{17}$~cm for $\Sdot = 3.7\EE{45}$~s$^{-1}$ 
and $\Sdot_{\rm ring} \ll \Sdot$. Because $r_{\rm b}$ is only a weak function
of ($\Sdot - \Sdot_{\rm ring}$), $r_{\rm b}$ could thus
be considerably smaller than the $2.5\EE{17}$~cm assumed by CD95. In a 1--D 
scenario, the region of hot shocked blue supergiant wind could therefore have
a thickness of as much as $\sim 3\EE{17}$~cm. The near-vacuum of
this region would not slow down the supernova ejecta appreciably, and an
extremely high initial velocity of the ejecta ($\gg 4\EE{4} \kms$; 
Chevalier 1998) would seem unnecessary. This agrees with explosion models of
the supernova which give maximum velocities of the ejecta $\sim 4\EE{4} \kms$ 
(Imshennik \& Nadyozhin 1989; Ensman \& Burrows 1992; Blinnikov et al. 1998).

The situation is different in a multidimensional scenario. While $r_{\rm b}$
may still be low, the radius of the contact discontinuity will retract in 
the equatorial plane due to the poleward flow of the shocked blue supergiant 
wind (Chevalier 1998). However, it should not retract too much to become 
incompatible with the evolution of the radio emission from the supernova, i.e.,
if we believe the maximum ejecta velocity is $\sim 4\EE{4} \kms$. 
To avoid such a retraction, the pressure in the equatorial plane of the 
H~II region should be lower too, compared to the 1--D case. It may be that the 
expected poleward gas flow in the H~II region (Chevalier 1998) is enough to
unload the pressure in the equatorial plane to allow for a radially extended 
shocked blue supergiant wind. Otherwise, we may have to invoke a lower 
photoevaporation rate than assumed in \S 2 to lower the pressure further.

The shapes of the termination shock, and the contact discontinuity between 
the blue and red supergiant winds, can only be studied in greater detail by
means of multidimensional hydrodynamical simulations (Chevalier 1998; ML98).
Detecting narrow lines from the H~II region, and how they vary in 
strength with time, would help to constrain the hydrodynamical models and to
obtain a more comprehensive picture of the presupernova nebula. This would aid
predictions of the ejecta/ring collision.

\section{Summary}
We have calculated the temperature and ionization of the H~II region created
by the progenitor of SN 1987A. We find that prior to the explosion,
the gas was probably cool, $T \lesssim 5000$~K, and had neutral helium. Using
this as input, we have calculated the ionization, temperature and line emission 
from the gas after it has been further heated and ionized by the radiation 
accompanying the supernova breakout. The models used for the breakout were 
500full1 and 500full2 of Ensman \& Burrows (1992). We find that the temperature 
of the gas stays relative constant at $\sim (0.6 - 1.0)\EE5$~K until the 
present epoch, and that the X-ray emission from the ejecta/wind interaction is 
unimportant for the evolution of temperature and ionization.

The slow recombination to O~III for the density inferred from the 
circumstellar X-rays, 150 a.m.u. cm$^{-1}$, appears to be inconsistent with
[O~III] observations at $t = 2500$~days. This argues that the [O~III]
emission was probably produced elsewhere than in the H~II region. This is 
also indicated by the limit on $n_{\rm H}$ we find from the early {\it IUE} 
observations of N~V~$\lambda1240$. Better lines to probe the density of the 
H~II region are instead N~V~$\lambda1240$ and O~VI~$\lambda1034$. They can be
observed using $\it HST$ and $\it FUSE$, respectively. Another important line 
is N~IV]~$\lambda$1486, whereas lines of lower ionization will only show
up if the density is a factor of $\sim 2$ higher than inferred from the X-rays.
The fluxes of all lines could start to decline well before day $\sim 5000$ 
due to the propagation of the blast wave.

The limits on density we find, together with the expected size of the H~II
region, put constraints on the nature of Sk~-69.202. We estimate that the upper
limit on its spectral type is not much earlier than B2~Ia, which is in accord
with the classification by Rousseau et al. (1978).

With the most likely inner and outer radii of the H~II region in the equatorial 
plane, $\sim 4.5\EE{17}$ and $\sim 5.9\EE{17}$~cm, respectively, we find that 
a 1--D approximation of the nebula could put the termination shock of the blue 
supergiant wind as close as $\lesssim 1.4\EE{17}$~cm to the star. This would 
mean an extended region of shocked blue supergiant wind, which would ease the
fast expansion of the ejecta. How this ties into a multidimensional scenario 
can only be explored by 2--D hydrodynamical simulations.

\acknowledgments
The author is grateful to Garrelt Mellema, Lifan Wang, Roger Chevalier, Claes
Fransson, Jason Pun and George Sonneborn for discussions, and to Dick McCray
for comments on the manuscript. This research was supported by the Swedish 
Natural Science Research Council, the Anna-Greta and Holger Crafoord Foundation,
and the Swedish Board of Space Research.

\clearpage

\figcaption[]{
Calculated light curves for a few important lines emitted by the H~II
region for three densities: $n_{\rm H} =$~60, 120 and $240 \cm3$.
The supernova breakout has been assumed to be similar to
that of the 500full1 model of Ensman \& Burrows (1992),
and the H~II region has been approximated by
a spherically symmetric shell with an outer radius of $5.9\EE{17}$~cm. In the
models marked by solid lines the inner radius of the H~II region is held fixed
at $4.5\EE{17}$~cm, whereas in the models marked by dashed lines, the increase
of the inner radius due to the propagation of the blast wave
has been included.  The distance to the supernova has been set to 50~kpc,
and no reddening has been included. The [O~III]~$\lambda$5007 observation is
from Wang et al. (1998), and the N~V~$\lambda$1240 observations are from
Sonneborn et al. (1997) assuming $E(\bv) = 0.16$ (see text). \label{fig1}}

\figcaption[]{
Same as in Fig.~1, but for the 500full2 model of Ensman \& Burrows (1992).
Note that the N~III], O~III] and [O~III] lines are notably weaker than in
Fig.~1. \label{fig2}}

\figcaption[]{
Calculated line ratios, relative to N~V~$\lambda1240$, for He~II~$\lambda1640$,
N~III]~$\lambda1750$, N~IV]~$\lambda1486$, O~III]~$\lambda1663$,
[O~III]~$\lambda5007$ and O~VI~$\lambda1034$ at $t = 4250$~days. The line
ratios are for four cases: the 500full1 burst with a reduced (see Fig. 1) H~II
region (solid lines),
500full1 with the full H~II region (dashed lines), 500full2 with a
reduced H~II region (long-dashed lines), and 500full2 with the full H~II region
(dashed-dotted lines). Note that while the absolute fluxes of the lines differ
significantly depending on whether the full, or reduced size of the H~II region
is included (see Figs.~1 and 2), the {\it ratios} of the lines are insensitive
to this. The effect of reddening has not been included. \label{fig3}}

\figcaption[]{
Calculated [O~III]~$\lambda5007$ flux at $t = 2500$ days as a function of
density for the models described in Figure~3. A distance of 50~kpc has been
assumed, but no reddening was included. The observed flux is from Wang et al.
(1998). Note the strong dependence of the flux on density. \label{fig4}}

\begin{figure}
\figurenum{1}
\epsscale{0.60}
\plotone{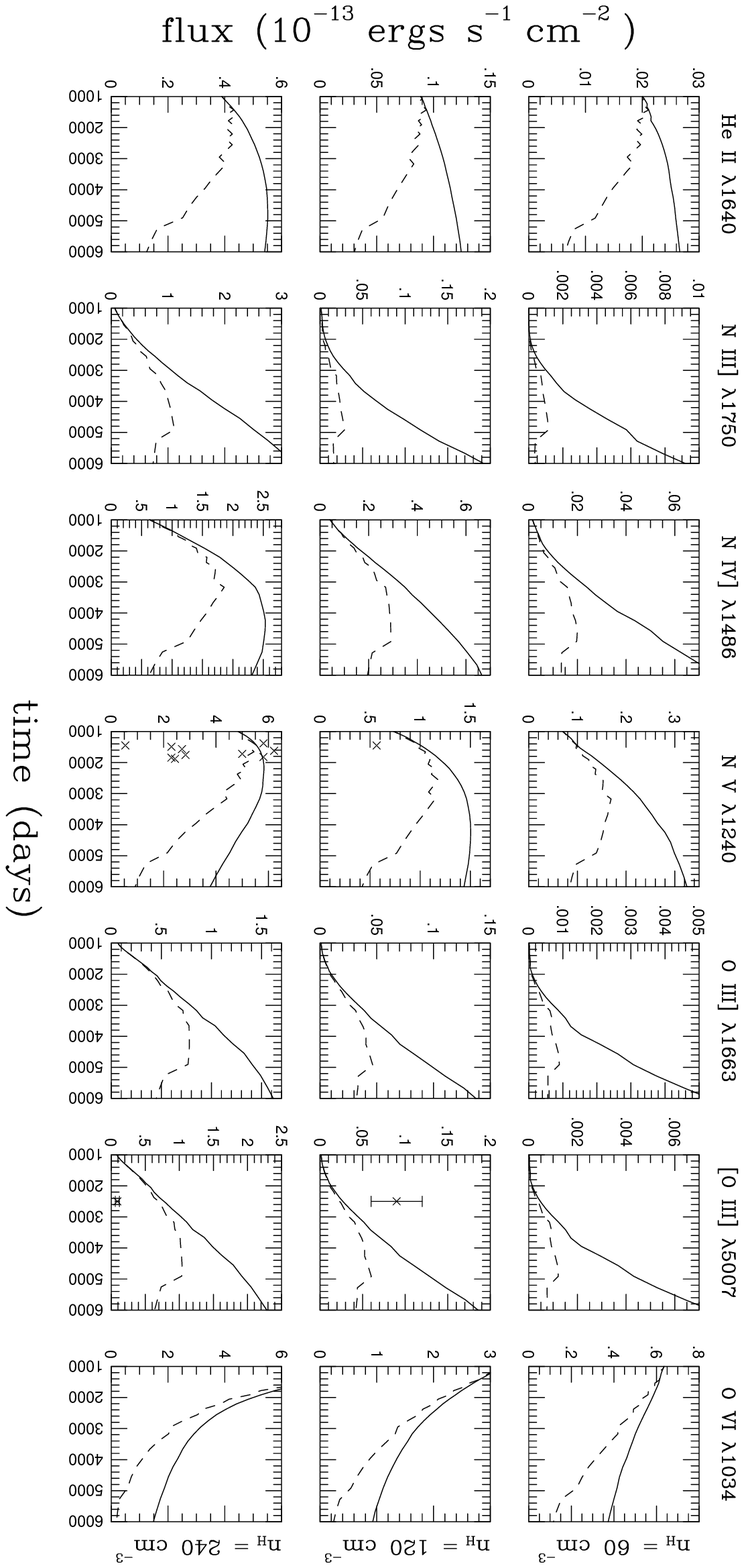}
\caption{}
\end{figure}

\begin{figure}
\figurenum{2}
\epsscale{0.60}
\plotone{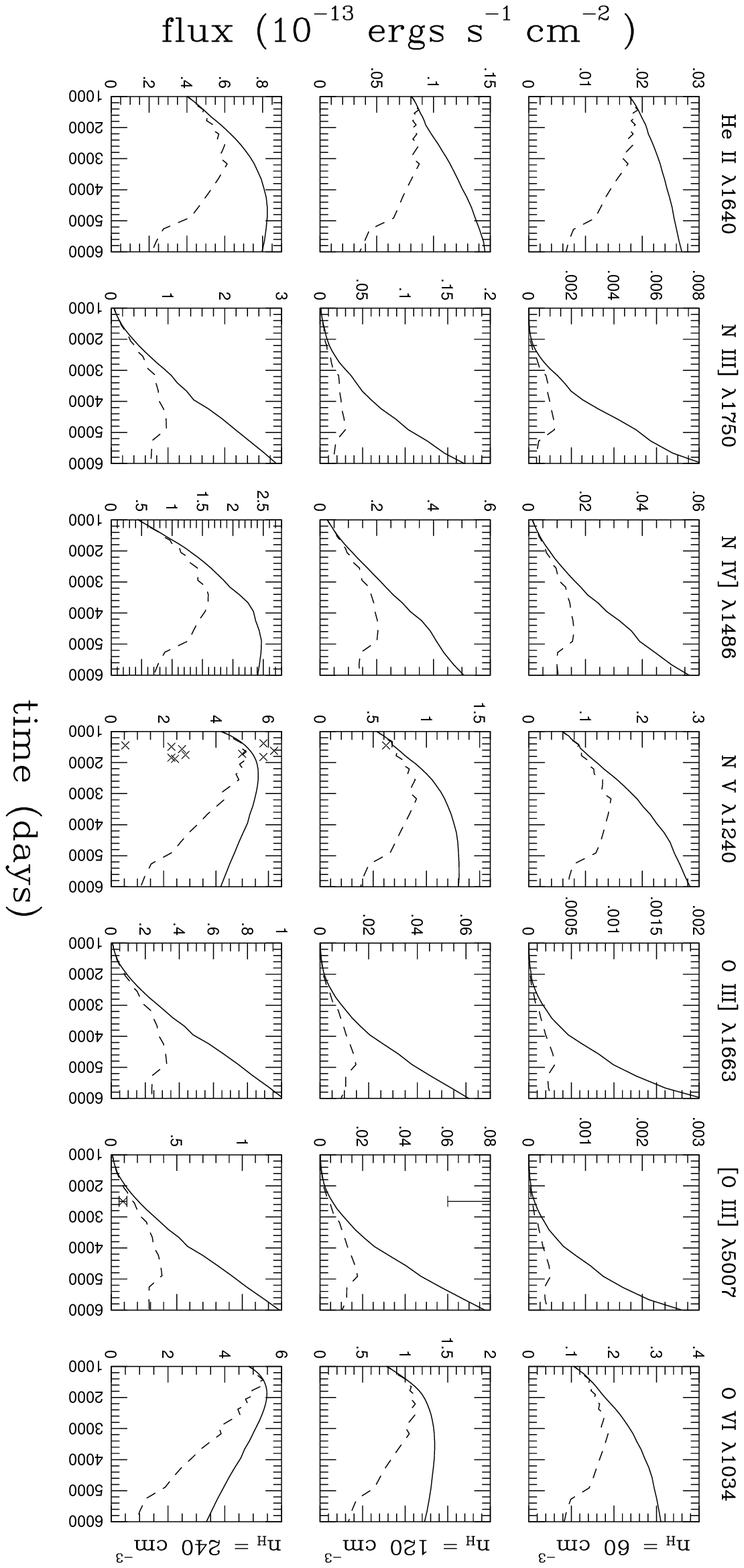}
\caption{}
\end{figure}

\begin{figure}
\figurenum{3}
\epsscale{0.25}
\plotone{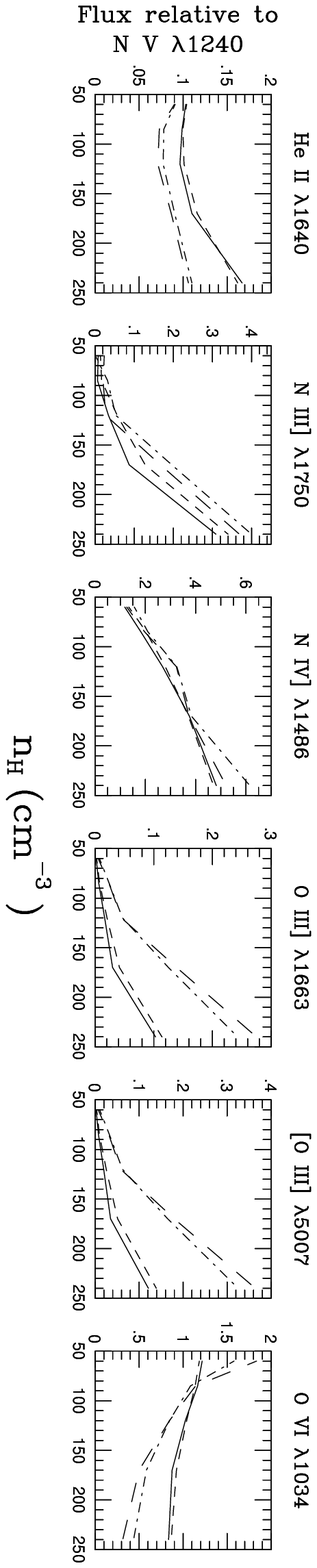}
\caption{}
\end{figure}

\begin{figure}
\figurenum{4}
\epsscale{0.70}
\plotone{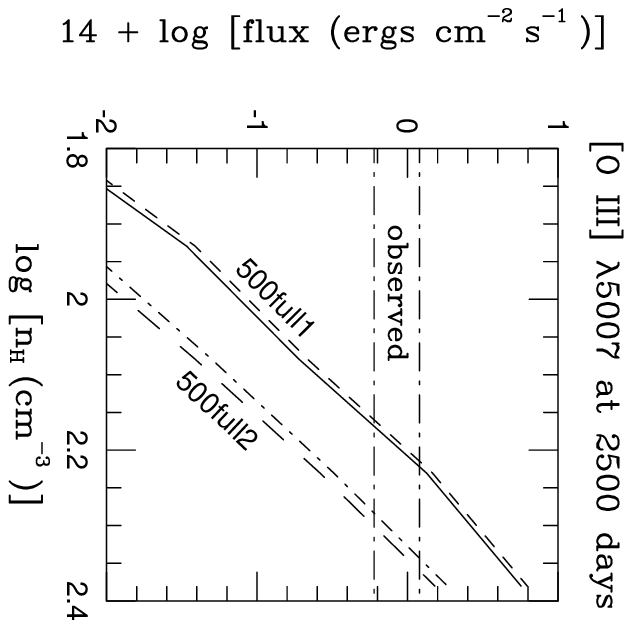}
\caption{}
\end{figure}

\end{document}